# Atomistic catalyst polarization stemming hydrogen generation from $CH_4$

Sanmei Wang[1], Yong Zhou[1], Chunyang Nie[1], Hengxin Fang[1], Biao Wang[1,2,*], Chang Q Sun[1,3,4,*]

## Abstract

As the extremely-sized nanocrystals and nanopores, an adatom (M) and atomic vacancy ($V^0$) exhibit extraordinary capability of catalysis with however little knowledge about the catalyst-reactant interfacial bonding dynamics. With the aid of density functional theory calculations, we examined the dehydrogenization of a single $CH_4$ molecule catalyzed using the Rh(111,100), W(110), Ru(0001) surfaces, and monolayer graphene, with and without M or $V^0$. It is uncovered in the following three components: (i) catalyst polarization due to atomic under- or hetero-coordination raises the valence band of the catalyst by bond contraction and atomistic dipolar ($M^P$) and vacancy dipolar ($V^0$) formation; (ii) reactant bond elongation by the interplay of the ($M^P/V^0$):H attraction and ($M^P/V^0$):⇔C repulsion with the ":" denoting the negative pole of the $M^P/V^0$; and (iii) reactant conversion, i.e., the scale of H-C elongation, the catalyst valence-band shift, the adsorption energy, and the catalytic activity are proportional to the charge quantity of the $M^P/V^0$ whose local electric field matters.

[1] Research Institute of Interdisciplinary Sciences (RISE) and School of Materials Science & Engineering, Dongguan University of Technology, Dongguan 523808, China
[2] Sino-French Institute of Nuclear Engineering and Technology, Sun Yat-Sen University, Zhuhai 519082, China
[3] Guangdong Provincial Key Laboratory of Extreme Conditions, Dongguan 523803, China
[4] School of EEE, Nanyang Technological University, Singapore 639798
Corresponding authors: wangbiao@mail.sysu.edu.cn (BW); 2022130@dgut.edu.cn; ecqsun@ntu.edu.sg (CQ)

TOC

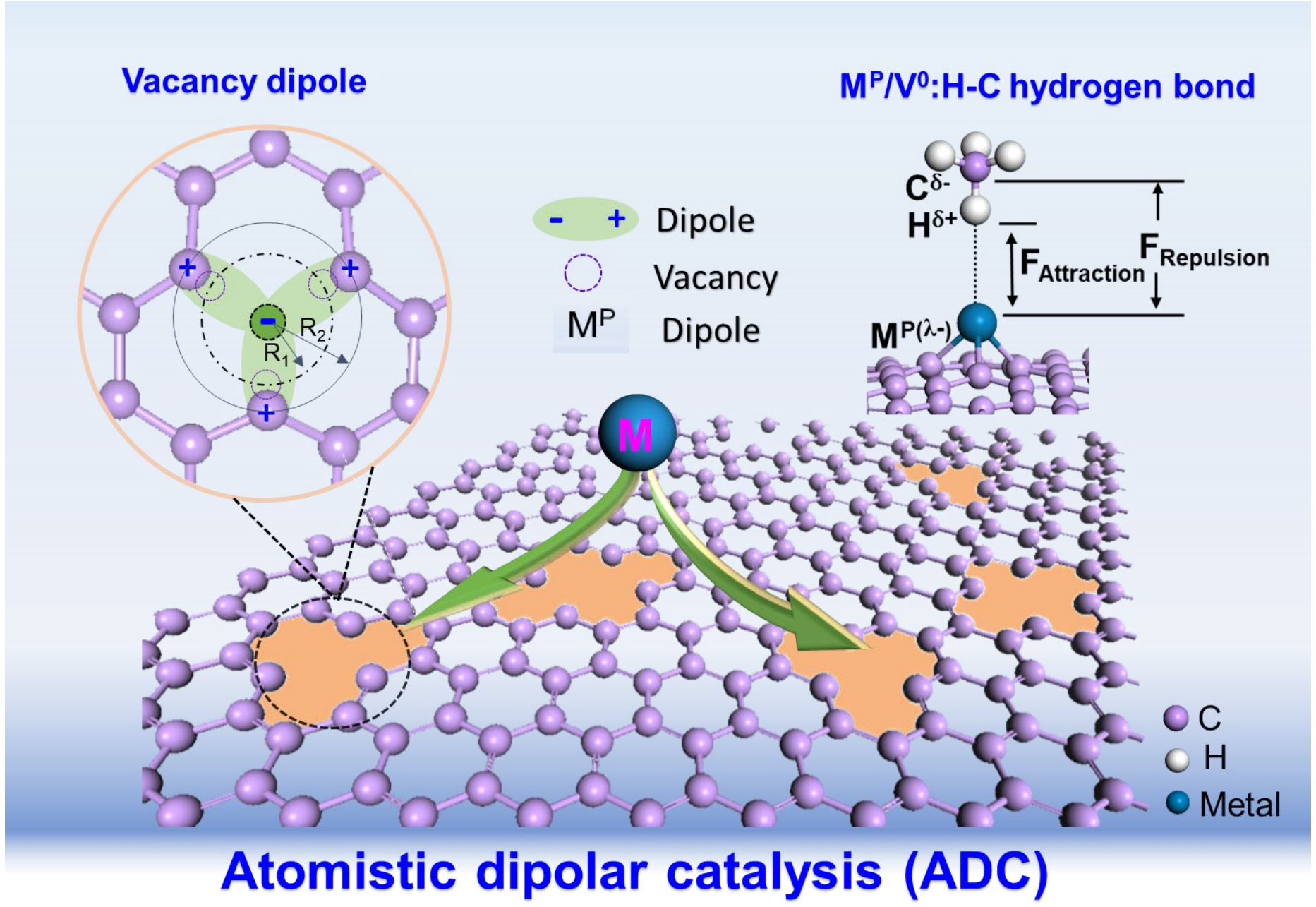

**The interplay of ($M^P/V^0$):H attraction and $V^0/M^P$:⇔C repulsion stretches the H-C bond whose degree of elongation depends to the local electric field of the $M^P/V^0$ atomistic dipolar catalyst (ADC).**

## 1. Introduction

As one of the key areas of chemical science and engineering, atomistic catalysis using undercoordinated adatoms, edge atoms, atomic vacancies [1-3], and hetero-coordinated atoms dispersed in ligand substrates [4] has attracted increasing attention due to their maximized efficiency in promoting reactions [5-8]. The distinction between various atomistic catalysts is their coordination environment which can be categorized into atomic undercoordination-induced atomic polarization and hetero-coordination interfacial polarization as an addition [9]. Nanostructured catalysts have tunable fractions of skin atoms of undercoordination for elemental species or the combination of under- and hetero-coordination. From the electronic configuration perspective, they are the same as polarization dominance. The dipole formed under varied coordination environments creates a strong electric field interacting with and activating the reactant molecules. Therefore, we may integrate the under- and hetero-coordinated catalysis into the same atomistic and nanoscale dipolar catalysis (ADC) category.

The undercoordinated atoms at terrace edges or facets account for ~70% of the total catalytic activity of a medium [9]. For instance, the even-undercoordinated adatoms on the Rh(111) surface favor $CH_4$ dehydrogenation more than atoms at the terrace edges [10-11]. A trace amount of Ru atoms added on the Cu(111) surface achieves a reaction rate 2.5 times higher than pure Cu nanoparticles in the light-driven $CH_4$ dry reforming with $CO_2$ [12]. The turnover frequency (TOF) of Ru adatoms on nitrogen-doped carbon for propane dehydrogenation is 3 times higher than that of the nanoparticle counterparts under the same conditions [13]. Additionally, the ensemble of Pt adatoms coordinated with oxygen atoms in MIL-101 achieves a TOF of 117 $h^{-1}$ in DMF at 32 bar pressure and 150 °C temperature for $CO_2$ hydrogenation into methanol, which is 5.6 times higher than that of Pt nanoparticle [14]. Ru single-atoms anchored on the RuNi alloy forming isolated alloy islands exhibiting 100% CO selectivity, over 55 times reverse water-gas shift (RWGS) rate than the alloys with Ru cluster sites [15].

The distorted 1T $MoS_2$ induced by lattice strain and Co-S bond formation between $Co_1$ and $MoS_2$ achieved Pt-like activity toward hydrogen evolution reaction and high long-term stability [16]. Heterogeneous Mo single-atom enzymes through the regulation of the coordination numbers (CNs) of the Mo sites exhibited superior and exclusive peroxidase-like activity [17]. The Fe-$S_1N_3$ configuration exhibits exceptional selectivity with CO Faradaic efficiency of 99.02 % and activity with TOF of 7804.34 $h^{-1}$, as well as remarkable stability during $CO_2$ reduction reaction [18]. Molecular dynamics calculations confirm lower activation barriers for the hydrogen evolution reaction (HER) at edge sites, with higher TOF of two to four orders of magnitude [19]. The $Cu_1/N_3$-BN [20] and $Rh_1/CeO_2$ [21] catalysis show an extremely high rate of $CH_4$ conversion and $N_2O$ dissociation. These findings evidence the significance of atomic undercoordination and hetero-coordination in atomistic catalysis which is promising in coping with energy and environmental crises.

Despite the promising achievements, there is still a strong desire to gain deeper insight into the origin and dynamics of the efficient ADC to devise catalysts as one wishes. Theories such as *d*-band center shift [22-23], nano-confinement [16], interfacial charge transfer [24-25], frontier orbitals [26], selective orbital coupling [19], and steric hindrance for pairing adatoms [27] have been developed and elegantly used. Among the known possible mechanisms, the *d*-band center theory correlates the catalysis efficiency directly to the extent of the *d*-band shift. According to this theory, reducing the catalyst sample size could raise the *d*-band towards the Fermi level. The *d*-band shift facilitates greater binding energy between the catalytic atoms and the adsorbed molecules. Hence, researchers tried to find means to enhance the catalytic efficiency by raising the *d*-band through impurity doping [28-33], atomic CN or manipulation by forming topological edges and atomic vacancies [33-34], mechanical straining [35], etc.

Reactant molecular adsorption is a crucially initial step in catalysis, such as $CH_4$ conversion catalyzed by $Rh_1/CeO_2$ [21], and $N_2O$ dissociation activated by $Cu_1/N_3$-BN [20]. The van der Waals interaction theory [36-38] explains the physical adsorption. However, questions may arise, such as how atomic undercoordination or hetero-coordination shifts the *d*-band of the catalyst, and what force activates and weakens the reactant bonds, which remain puzzling.

In this work, we combined bond order-length-strength correlation and nonbonding electron polarization (BOLS-NEP) [9, 39] and hydrogen bond cooperativity and polarizability (HBCP) theory [40-41], as well as density functional theory (DFT) calculations to examine the dehydrogenation of $CH_4$ onto a metal adatoms ($M_1$) and vacancies ($V_0$) on the homogeneous ligand substrates (M = Rh, Ru, and W) and heterogeneous graphene. Results show that atomic undercoordination shortens the local bond, densifies and entraps core and bonding electrons, polarizes the valence electrons of the metal adatoms ($M_1$) and vacancies ($V_0$) to make them dipoles ($M^P/V^0$) with excessive charge quantities. When $CH_4$ adsorption on surface, the atomistic dipolar $M^P/V^0$ interacts with $CH_4$ molecule by forming the dipolar-substituted hydrogen bond ($M^P/V^0$:H-C) with the ":" denoting the negative pole of the $M^P/V^0$. The interplay of $M^P/V^0$:H attraction and $M^P/V^0$:⇔C repulsion lengthens the H-C bond of $CH_4$, activating $CH_4$ toward dissociation.

## 2. Principles and predictions

### 2.1 BOLS-NEP and HBCP regulations

The BOLS-NEP [9, 39] and HBCP [41-42] notions (Fig. A1 in Appendix 1) are expressed as follows:

$$\begin{cases} C_z & = d_z/d_b = 2\Big/\left\{1+exp\left[(12-z)/(8z)\right]\right\} \\ C_z^{-m} & = E_z/E_b \end{cases} \quad \text{(a, BOLS-NEP)}$$

$$\begin{cases} d_{OO} & = 2.6950\rho^{-1/3} \\ d_L/d_{L0} & = 2\Big/\left\{1+exp\left[(d_H-d_{H0})/0.2428\right]\right\} \end{cases} \quad \text{(b, HBCP)}$$

(1)

Where $C_z$ is the bond contraction coefficient that depends only on the atomic-CN($z$), regardless of the chemical composition of a substance, $d$ is the bond length and $E$ is the bond energy, subscripts b and z denote respectively the fully- and the z-coordinated atoms. The $m$ is the bond nature index that correlates the energy to the bond length $E \propto d^{-m}$. The ρ is the mass density of water, $d_{OO}$ denotes the O—O distance of water with $d_L$ being the O:H length and $d_H$ the H-O length, projecting along the O—O line. The $\rho = 1.0$ g/cm$^3$, $d_{L0} = 1.6946$, and $d_{H0} = 1.0004$ Å for the $H_2O$:4$H_2O$ motif at 277 K and the ambient pressure, are taken as references.

The BOLS-NEP notion indicates that atomic or molecular undercoordination shortens and strengthens the bonds between undercoordinated atoms, associated with the core and bonding electron self-entrapment and nonbonding electron polarization (Fig. S1a). Bond contraction densifies the local electrons and bond strengthening deepens the local potential well that entraps the local electrons. The locally and densely entrapped core and bonding electrons polarize the valence electrons of the even undercoordinated atoms, making them dipoles ($M^P/V^0$) with excessive charge quantities. The excessive dipolar charge forms a local electric field. Fig.1 illustrates schematically the formation of the dipolar atomic vacancy ($V^0$) and dipolar adatom ($M^P$). Removing an atom from the graphene leaves behind a shell of denser charge and energy surrounding the vacancy because of the bond contraction. The denser energy enhances the local mechanical strength, facilitating energy storage and nanoporous and defect mechanics [43-44]. The denser charge polarizes the vacant edge atoms, making them centralized oppositely-coupled dipoles and the Dirac-Fermion resonant peak at the Fermi energy ($E_F$) [45-46], which contributes to the presently-addressed atomistic dipolar catalysis in which the dipolar $V^0$ further polarizes the metal adatom on top of the vacancy into an $M^P$ dipole.

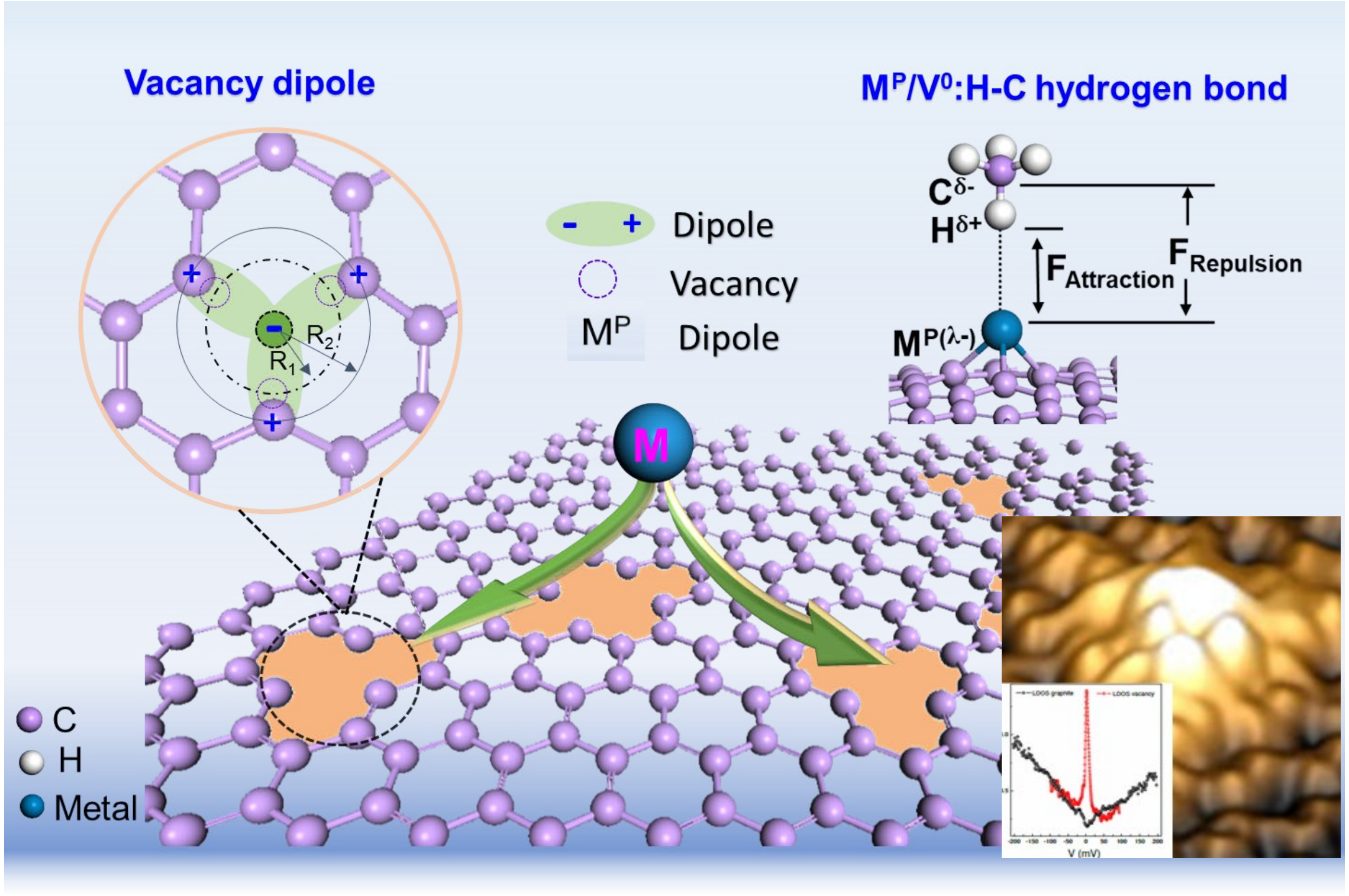

Fig. 1 Schematic diagram of formation of vacancy dipole ($V^0$) and dipolar-substituted hydrogen bond ($M^P/V^0$:H-C). Vacancy formation dislocates the edge atoms radially away from the vacancy and polarizes them into dipoles pointing to the center of the vacancy, according to the BOLS-NEP regulation [9]. The inset shows the atomic vacancy resolved

Dirac-Fermion and STM protrusion on the graphite surface [46] and monolayer graphene [45].

The HBCP mechanism for water indicates that the three-body coupling O:H-O bond relaxes cooperatively in its segmental length and energy when perturbed because of the interplay of the O:H attraction and O:⇔:O repulsive coupling [47]. The O—O distance changes by shortening one segment lengthening the other, and the O:H nonbond always relaxes more in scale than the H-O in contrasting directions. This premise has led to a quantitative resolution to anomalies of water and ice such as ice floating, ice friction, ice regulation, warm water fast cooling, etc. Progress has formed the subject of a recent treatise [41]. In this work, we will show that the HBCP regulation for water is applicable to the $CH_4$ dehydrogenation reaction by extending the O:H-O to the dipolar-substituted hydrogen bond ($M^P/V^0$:H-C).

### 2.2 Dipolar-substituted coupling $M^P/V^0$:H-C bond

When $CH_4$ is adsorbed on the undercoordinated surface, dipolar atomic vacancy ($V^0$) and dipolar adatom ($M^P$), induced by bond contraction between under-coordination atoms, will polarize the $CH_4$ molecules, forming dipolar-substituted $M^P/V^0$:H-C bond, as illustrated in the Fig. 1 inset. The ":" signifies the negative pole of the $M^P/V^0$. The variation of segment length of $M^P/V^0$:H-C follows the HBCP rule. The combination of $M^P/V^0$:H attraction and $M^P/V^0$:⇔C repulsion stretches the H-C bond toward dissociation. The extent of H-C elongation and the catalytic efficiency are proportional to the $M^P/V^0$ charge quantity that increases with the degree of atomic undercoordination. The dipolar-substituted $M^P/V^0$:H-C bond is anticipated to relax cooperatively in its segmental length and energy. Depending on its charge quantity, the electric field of the $M^P/V^0$ polarizes the reactant H-C bond, leading to the $M^P$:H distance contraction and H-C expansion.

## 3. Results and discussion

### 3.1 Valence band shift by atomic undercoordination

We firstly examined the effect of bond contraction on the valence band shift of a metal adatoms ($M_1$) and vacancies ($V_0$) on the homogeneous ligand substrates (M = Rh, Ru, and W) is first examined. Fig. 2 shows the structural configurations and the differential local density-of-state of the valence band (ΔLDOS) for the considered Rh(111;100), Ru(0001), and W(110) surfaces with and without $M^P/V^0$. The distance between atoms denoted 1 and 2 is showed in Fig.3a, c, and e. We can see that the "1-2" distance contracts compared to the bulk, though the DFT-derived strains are lower than those obtained by XPS refinement (Table S1) [9].

According to BOLS-NEP notion, the contraction of "1-2" bond will densify the local electrons and bond strengthening deepens the local potential well that entraps the local electrons. The locally and densely entrapped core and bonding electrons polarize the valence electrons of the even undercoordinated atoms. Based on this prediction, we further investigate the ΔLDOS shifts, as shown in Fig. 2b, d, and f. The valence LDOS transition from the bulk to the adatom is obtained by subtracting the LDOS of the bulk standard from the one with an $M^P/V^0$ after the LDOS is area normalized. An LDOS peak integral is proportional to the charge of the adatom Δq (in fraction) transferring from bulk (spectral valley). Therefore, one can obtain the valance band (VB) center shift ($\Delta E_{VB}$) and the fraction of electron transferring (Δq) to the atomistic dipole by integrating the LDOS using equations given in Appendix 2.

Table S1 shows the VB-band shift or the atomic-CN variation through local bond contraction and dipole formation, consistently demonstrating atomic-CN dependence. For the dipolar Rh ($Rh^P$) on the (111) surface, the values of $\Delta E_{VB}$ and Δq follow the order of $Rh_1 > Rh_3^0 > Rh_1^0 >$ Rh(111). For the dipolar $Rh^P$ on the (100) surface, the values of $\Delta E_{VB}$ and Δq follow the order of $Rh_1 > Rh_1^0 >$ Rh(100). For dipolar $Ru^P$, the values of Δq follow the order of $Ru_1 > Ru_3^0 > Ru_1^0 >$ Ru(0001). The values of $\Delta E_{VB}$ for $W^P$ on the (110) surface follow the order of $W_1 > W_1^0 >$ W(110).

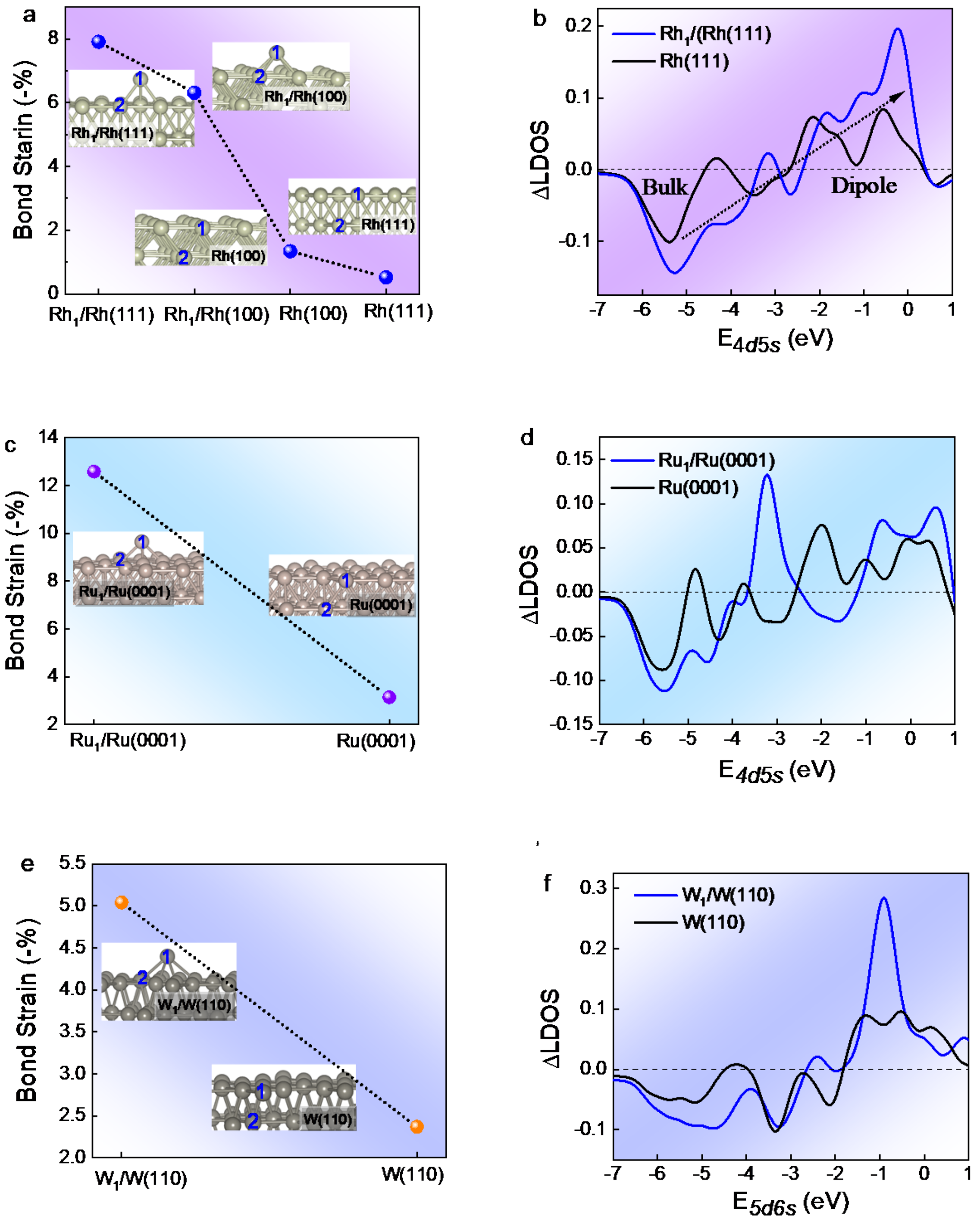

Fig.2 Atomic-undercoordination-resolved bond contraction and VB-band center upward shift. Computationally derived bond strain and the differential valence ΔLDOS for the (a, b) Rh(111), (c, d) Ru(0001), and (e, f) W(110) surfaces with and without adatoms. The ΔLDOS for the $Rh_1$/Rh(100) and Rh(100) are similar but less significant than that for the $Rh_1$/Rh(111) and Rh(111). The Fermi level is set to 0 eV.

## 3.2 Dipolar-substituted coupling $M^P/V^0$:H-C bond

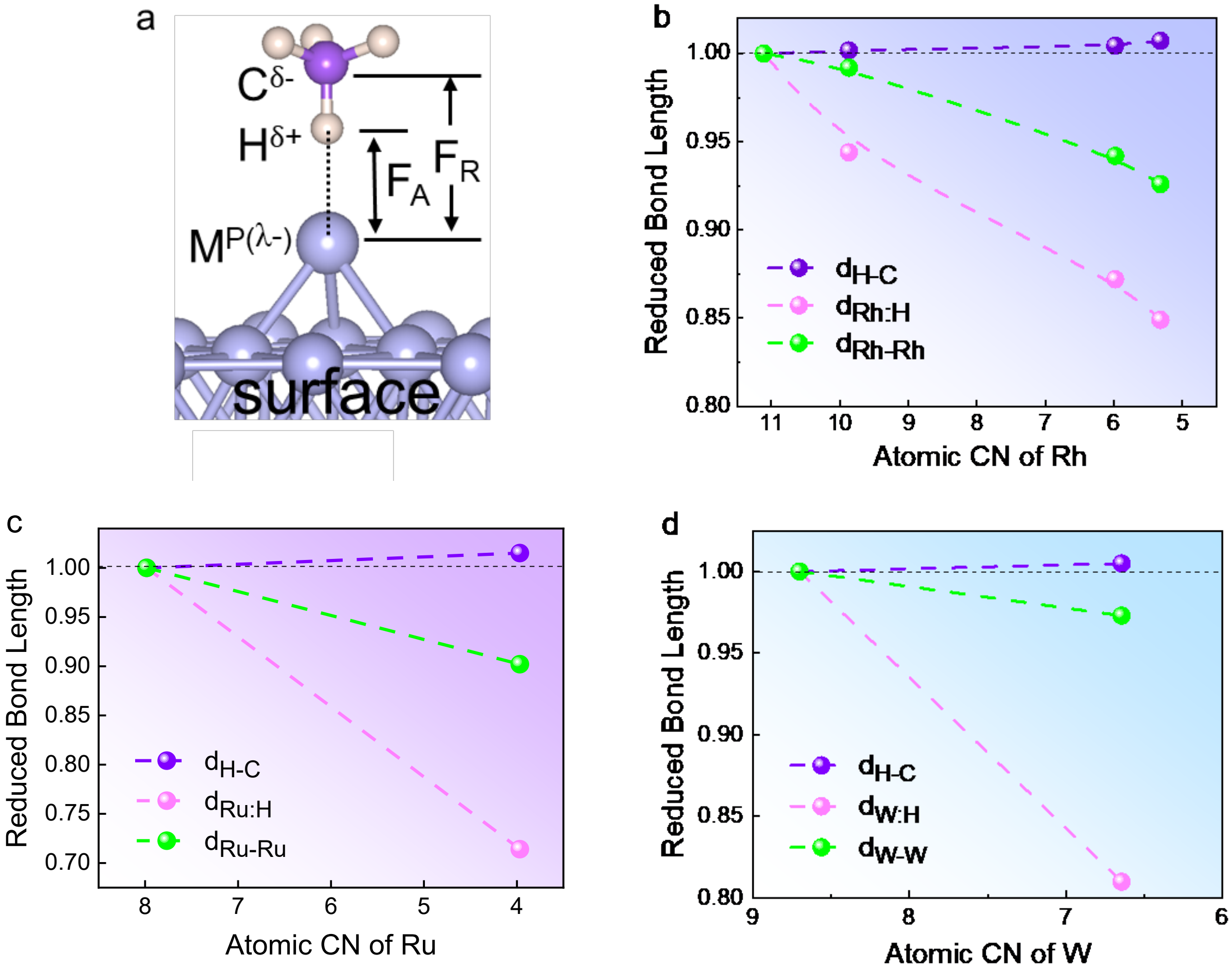

Fig. 3 Cooperative relaxation of the dipolar-substituted $M^P$:H-C bond. (a) The interplay of $M^P$:H attraction ($F_A$) and $M^P$:⇔C repulsion ($F_R$) stretches the H-C toward dissociation. The M-$M^P$:H-C segmental length cooperativity for $CH_4$ adsorption on the (b) Rh, (c) Ru, and (d) W surfaces with and without adatoms.

When $CH_4$ is adsorbed on the surface, the dipolar atomic vacancy and dipolar adatom, induced by bond contraction between under-coordination atoms, will polarize $CH_4$ molecules forming the dipolar-substituted $M^P/V^0$:H-C hydrogen bond. Fig. 3a illustrates the M-$M^P$:H-C configuration for the catalyst-reactant interaction. $CH_4$ may interact with the $M^P/V^0$ through configurations involving one, two, or three of its hybridized orbitals. The atomistic dipole $M^P$ or vacancy $V^0$ replaces the X in the X:H-C coupling hydrogen bond to form the $(M^P/V^0)^{\lambda-}$:$H^{\delta+}$-$C^{\delta-}$ with the adsorbing H−C molecular dipole. The value λ varies with the extent of catalytic atomic undercoordination, and the δ with the electronegativity difference between the H−C dipolar constituents. The $(M^P/V^0)^{\lambda-}$:$H^{\delta+}$-$C^{\delta-}$ bond adheres to the force combination criterion and follows the HBCP regulation [41]. One may omit the valence values and signs for convenient discussion. Applying an electric field in a proper direction could raise the λ and δ values and enhance the catalytic efficiency by strengthening the ($M^P/V^0$):H attraction and the ($M^P/V^0$):⇔C repulsion, which further lengthens and weakens the H−C toward dissociation.

Fig. 3b-d and Table S1 show the segmental length cooperativity of the M-$M^P/V^0$:H-C system as a function of atomic-CN. The lower the $M^P$ coordination, the shorter the distance and the greater the charge transfer from the positive pole of the $M^P$ to the negative. The H-C elongation is the key indicator of catalytic efficiency. From this perspective, the efficiency is in the $Rh_1 > Rh_3^0 > Rh_1^0 >$ Rh(111) for the dipolar $Rh^P$ on the (111) surface, as illustrated in Table S1. The Ru and W are also in the same significance order though the $W_3^0$ is not considered as it is hard to form such a vacancy on the W(110) surface. The Rh(100) surface is slightly strange due to possible geometric configuration.

Additionally, we investigated the local electric field induced by $M^P$ dipole using $Rh_1$/Rh(111), $Ru_1$/Ru(0001), and $W_1$/W(110) as illustrative cases. The electric field strength was acquired by calculating the gradient of the electrostatic potential. Electric field strength at a point can be defined as the negative of the potential gradient at that point, $E = -\text{grad}\ V$. Fig. 4a-c shows that for $Rh_1$/Rh(111), $Ru_1$/Ru(0001), and $W_1$/W(110), the local electric fields strength at distances $M^p$ of 0.61, 0.63, and 0.77 Å are -1.83, -2.11, and -0.91 V/Å, respectively. From Table S1, we observe that the H-C bond elongations parallel these intensities, being 1.110, 1.116, and 1.108 Å, respectively. This correlation suggests that an increase in electric field strength leads to a longer H-C bond.

To further verify the polarization and elongation of the C-H bond owing to the electric field, we apply external electric fields of -1.83, -2.11, and -0.91 V/Å to an isolated $CH_4$ molecule. Consequently, the C-H bond lengths expand from 1.097 Å to 1.102, 1.103, and 1.098 Å, respectively. These findings confirm that the stretching of the C-H bond of $CH_4$ on metal surfaces is indeed caused by the local electric field formed by the atomistic dipoles. Furthermore, we observe that at equal electric field strength, the extent of C-H bond elongation of isolated $CH_4$ molecule induced by an external electric field is lesser that of adsorbed $CH_4$ on surface caused by the dipole-generated electric field. This could potentially stem from the fact that when $CH_4$ physically adheres to the surface, it's not just influenced by the $M^P$ dipole's electric field but also by the electric field emanating from other atoms on the flat surface.

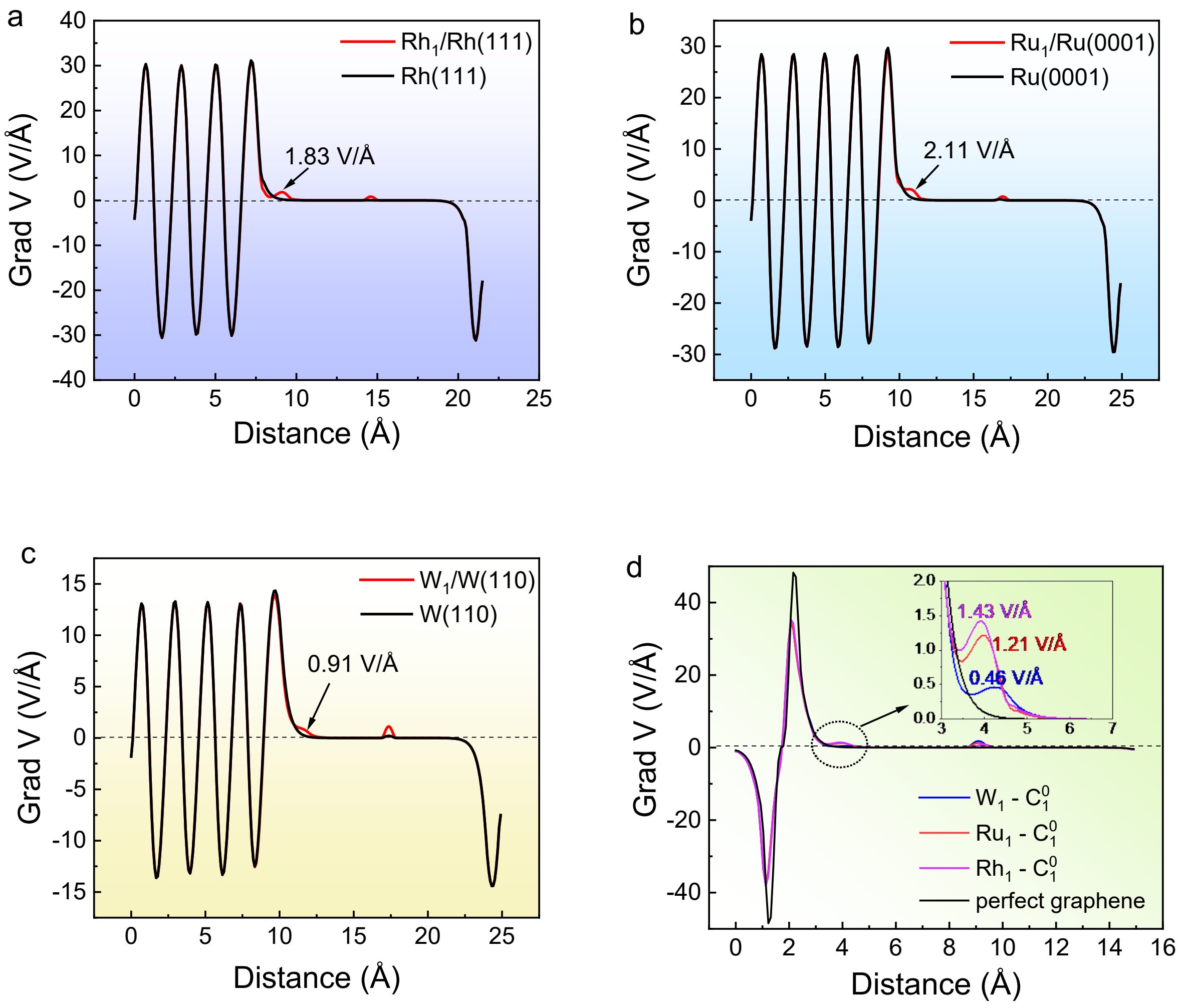

Fig. 4 The gradient of the electrostatic potential (grad V) at different positions on (a) $Rh_1/Rh(111)$, (b) $Ru_1/Ru(0001)$, (c) $W_1/W(110)$, and (d) $M_1$- $C_1^0$.

Next, we explored the situation of $CH_4$ adsorption on a metal atom supported on heterogeneous graphene. When C atom vacancies appear in graphene, the edge atoms of vacancy dislocate radially away from the center of the vacancy while forming dipoles pointing to the center of the vacancy. The dipolar $V^0$ polarizes the M adatom located on top of the vacancy into an $M^P$ dipole. When $CH_4$ adsorption on surface, the $M^P$ dipole forms the $C^0$-$M^P$:H-C coupling hydrogen bond with the H-C bond of $CH_4$, as illustrated in Fig. 1

Fig. 5 and Table S2 show the effect of atomic vacancies ($C^0$) on the $C^0$-$M^P$:H-C cooperative relaxation and the valence ΔLDOS shift of the $M^P$ that sits on the single carbon atomic vacancy ($C^0_1$) and two-atomic vacancy ($C^0_2$) of graphene. Carbon atoms in graphene undergo the $sp^2$-orbital hybridization with each atom having an unpaired lone electron capable of polarizing the metallic adatom. Calculations revealed that graphene with and without atomic vacancy is in the G > $C_2^0$ > $C_1^0$ significance order for the H-C elongation in the $V^0$-$M^P$:H-C coupling system.

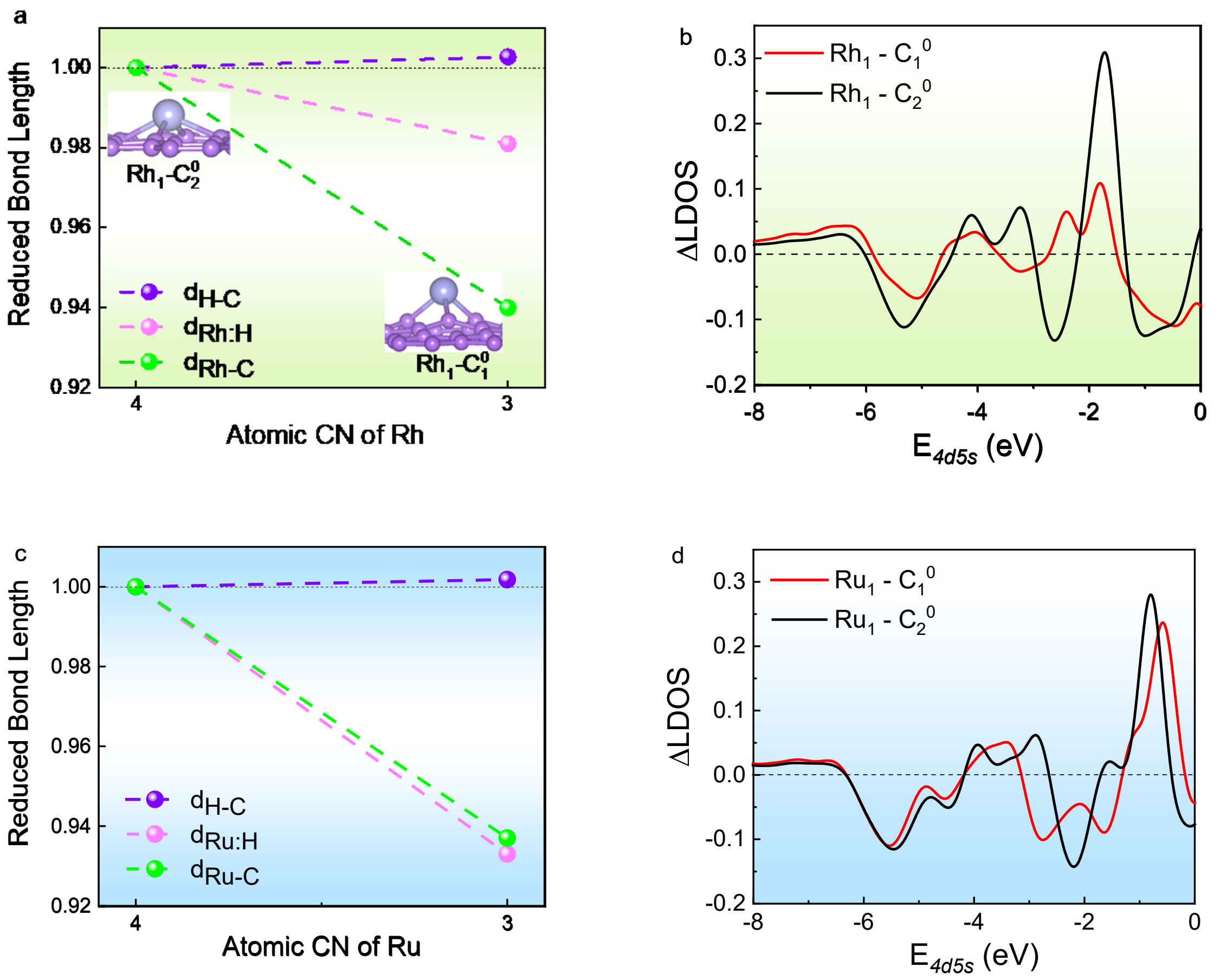

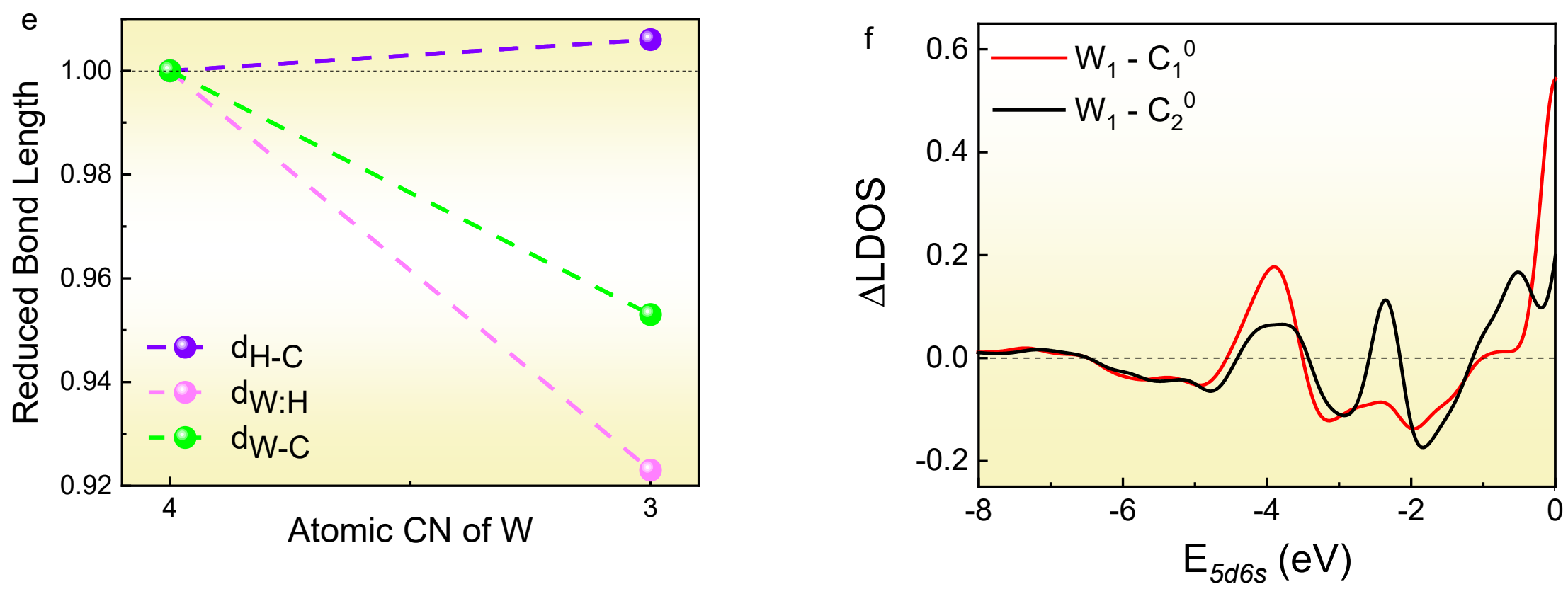

Fig. 5 Cooperative relaxation of the dipolar-substituted $M^P$:H-C bond and the valance LDOS shift for the M adatoms located on $C^0$ atomic vacancies. The $C^0$-$M^P$:H-C segmental length cooperativity for $CH_4$ adsorption on graphene with (a) Rh, (c) Ru, and (e) W atoms doped on $C_1^0$ and $C_2^0$ vacancies and the (b-f) corresponding ΔLDOS.

We use $Rh_1$-$C_1^0$, $Ru_1$-$C_1^0$, and $W_1$-$C_1^0$ as examples to explore the local electric field formed by $M^p$ dipoles. Fig. 4d shows that for $Rh_1$-$C_1^0$, $Ru_1$-$C_1^0$, and $W_1$-$C_1^0$, the local electric fields strength at distances $M^p$ of 0.65, 0.7, and 0.9 Å are -1.43, -1.21, and -0.46 V/Å, respectively. To further verify the polarization and elongation of the C-H bond owing to the electric field, we apply the same electric field to an isolated $CH_4$ molecule. Upon application of the external electric field (-1.43, -1.21, and -0.46 V/Å), the C-H bond lengths increased slightly from 1.097 Å to 1.100, 1.099, and 1.098 Å, respectively. These findings confirm that the local electric field of the atomistic dipole provides the unique force stretching the reactant C-H bond.

Fig. 6 shows the cooperative relaxation of the M/C-$M^P$/$V^0$:H-C bonding system and the correlation between the net charge gain and the VB shift of the metallic catalyst on metals and graphene. The extent of the H-C bond elongation varies with the $M^P$/$V^0$:$CH_4$ configuration. The net charge gains Δq of the catalytic atom is proportional to the valence band center shift and the extent of H-C elongation for both M/C-$M^P$/$V^0$:H-C scenarios. The ε exists not for metal on graphene. The results evidence that M/C-$M^P$/$V^0$:H-C bonding system follows the HBCP regulation.

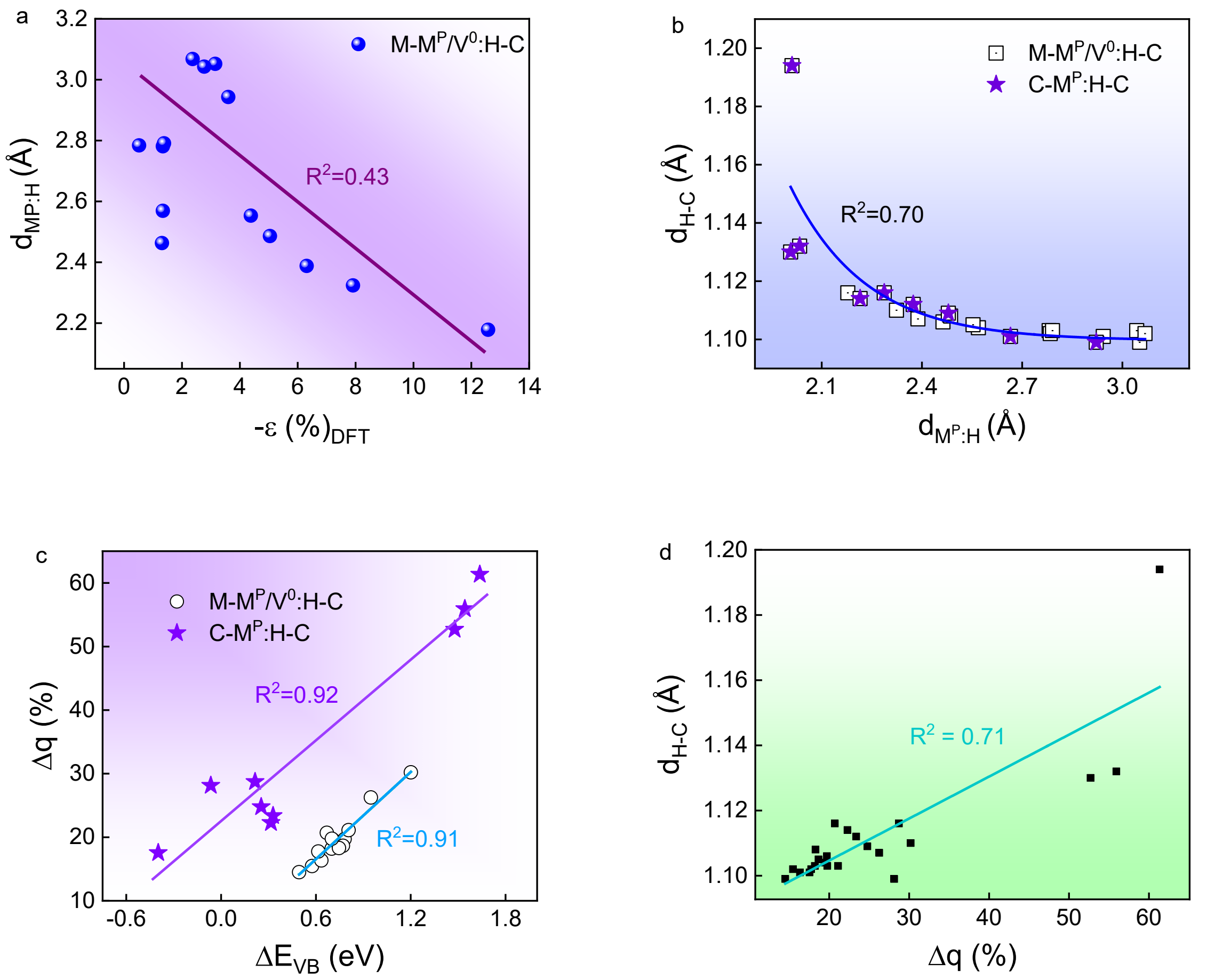

Fig. 6 Cooperative relaxation of the M/C-M$^P$/V$^0$:H-C bonding system and correlation between catalyst charge gain and VB shift. (a, b) The segmental lengths correlation and the nearly linear dependence of the charge quantity gain on (c) the valence band shift of the atomistic metallic dipolar catalysts on metal and graphene and the (d) H-C bond length. Provided are reference guidelines.

## 4. Pathways of $CH_4$ dehydrogenation by Rh adatom

Next, attention is given to the atomic-CN dependence of the adsorption energy and dehydrogenation pathways for the peculiarly coordinated Rh catalyst. The adsorption energy ($E_{ads}$) of $CH_4$ is calculated as $E_{ads} = E_{(surf+CH4)} - (E_{surf} + E_{CH4})$. $E_{surf}$ and $E_{(surf+CH4)}$ represent the system energies before and after $CH_4$ adsorption. $E_{CH4}$ is the energy of $CH_4$ molecules. The calculated values of $E_{ads}$ on $Rh_1$/Rh(111;100) and Rh(100;111) surfaces were -0.35, -0.27, -0.22, and -0.21 eV, respectively, showing the atomic-CN dependence. Fig. 7a-c illustrates a linear correlation between these adsorption energies and factors such as charge gain, H-C bond length, VB-center, and d-band center. An increase in charge quantity and an upward shift of both VB-center and *d*-band center promote $CH_4$ adsorption (Fig. 7a,b). Furthermore, stronger adsorption energy leads to increased H-C bond elongation, enhancing $CH_4$ activation efficacy (Fig. 7c).

Fig. 7d shows the possible dehydrogenation pathways and the corresponding energy profiles. The free energy barriers for the H-C bond cleavage of $CH_4$ to form $*CH_3$ and *H species on $Rh_1$/Rh(111;100), and Rh(100; 111) surfaces are calculated as 0.08, 0.55, 0.53, and 0.15 eV, respectively. Among these four structures, $Rh_1$/Rh(111) possesses the lowest energy barrier for the H-C bond cleavage, highly capable of breaking the $CH_4$ bond than other coordination configurations. The stronger cleavage ability of the $Rh_1$/Rh(111) is attributed to the local electric field generated by the $M^P$ dipole on $Rh_1$/Rh(111) weakens and elongates the $CH_4$ bond more effectively than other configurations during the physical adsorption of $CH_4$.

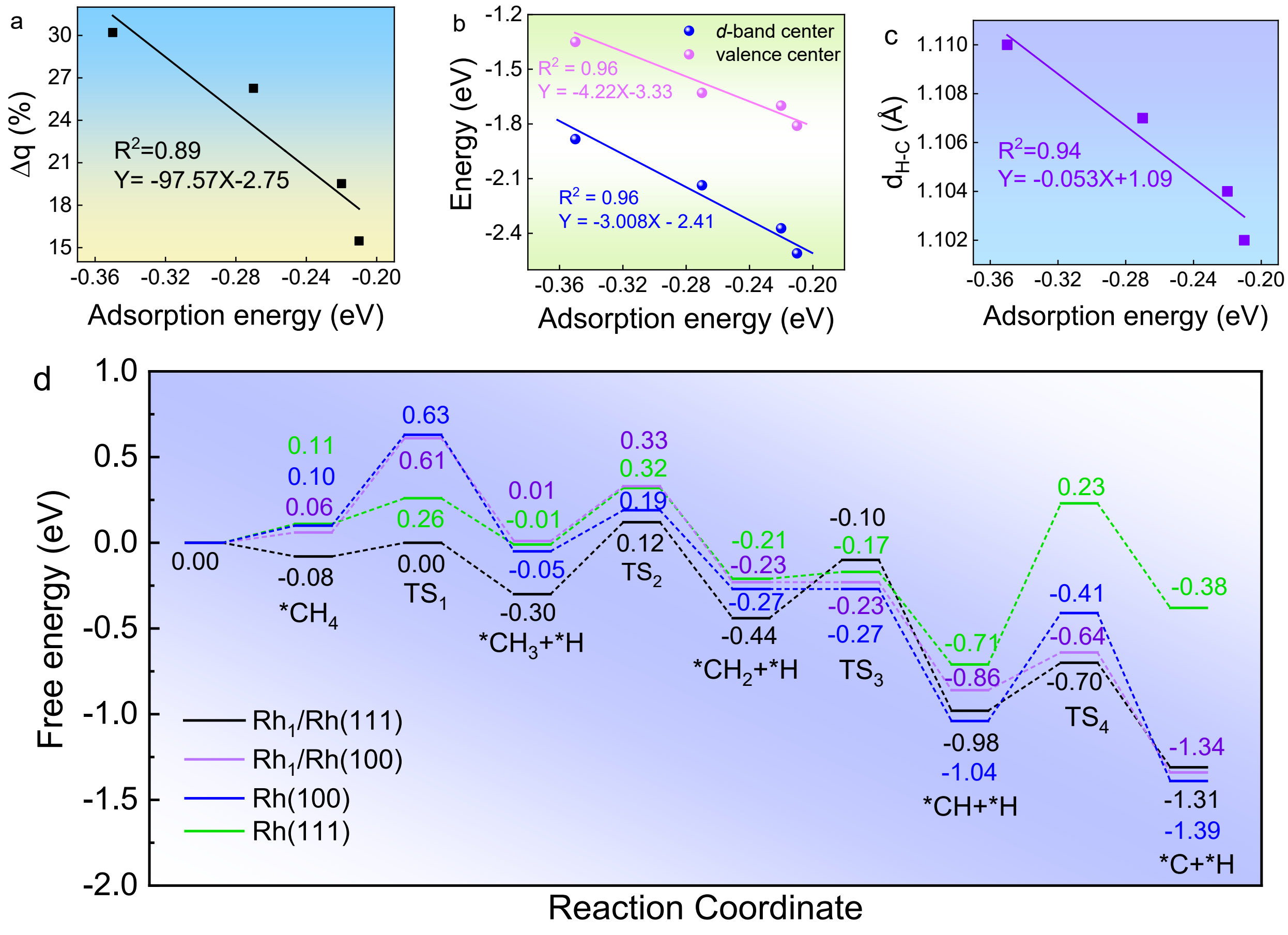

Fig. 7 The correlation between adsorption-energy and charge gain, VB-center, d-band center, and H-C length, and atomic undercoordination resolved dehydrogenization barriers. The linear dependence of the adsorption-energy on (a) charge gain, (b) VB-center, and d-band center, and (c) H-C length of adsorbed $CH_4$. (c) The free energy profiles for $CH_4$ dehydrogenation over $Rh_1$/Rh(111;100) and Rh(111;100) surfaces.

The *$CH_3$ species formed by the cleavage of the H-C bond in $CH_4$ are unstable, and may further combination or dehydrogenation. For $Rh_1$/Rh(111), the free energy barrier for the combination of two *$CH_3$ species is 1.24 eV, which is higher than that for the dehydrogenation of *$CH_3$ (0.42 eV) (Fig. 7d), indicating that the *$CH_3$ species on $Rh_1$/Rh(111) tend to undergo dehydrogenation to form *$CH_2$ species. A similar phenomenon is observed for $Rh_1$/Rh(100), Rh(100), and Rh(111). When *$CH_2$ species form, they may directly combine to or further dehydrogenate to generate *CH species. For the four structure configurations, the coupling of two *$CH_2$ species shows higher energy barriers than the generation of *CH species, suggesting that *$CH_2$ species tend to dehydrogenate to form *CH species. Furthermore, we evaluate the possibility of coupling of *CH species and the dehydrogenation of *CH species by comparing their free energy barriers. Results show that the coupling of *CH species is blocked by its higher free energy barriers compared to *CH species dehydrogenation on four structures. Thus, the $CH_4$ on all structures is more favorable for successive dehydrogenation than the C-C coupling.

The rate-determining steps (RDS) for $CH_4$ dehydrogenation to C and 4H on the $Rh_1$/Rh(111; 100) and Rh(100;111) system involve the dissociation of *$CH_3$, *$CH_4$, *CH, and *$CH_4$, respectively. The free energy barriers of RDS are 0.42, 0.55, 0.63, and 0.94 eV, respectively (Fig. 8a). Among the four structures, $Rh_1$/Rh(111) possesses the lowest RDS energy barriers, indicating its higher catalytic activity. To comprehend why $Rh_1$/Rh(111) outperforms three other structures in catalytic activity, we investigate the correlation between RDS energy barriers and factors such as charge gain, H-C length, VB-center shift. Fig. 8b, c reveal that $Rh_1$/Rh(111) stands out with more significant charge accumulation and a larger VB-center shift. During CH4's physical adsorption, $Rh_1$/Rh(111) more effectively activates $CH_4$ by elongating the H-C bond (Fig. 8d). Furthermore, the liner correlation in Fig. 8b-d indicate that charge gain, VB center shift, and H-C length can serves as a reliable descriptor of catalytic activation for accelerating the discovery of highly efficient catalysts for $CH_4$ dehydrogenation.

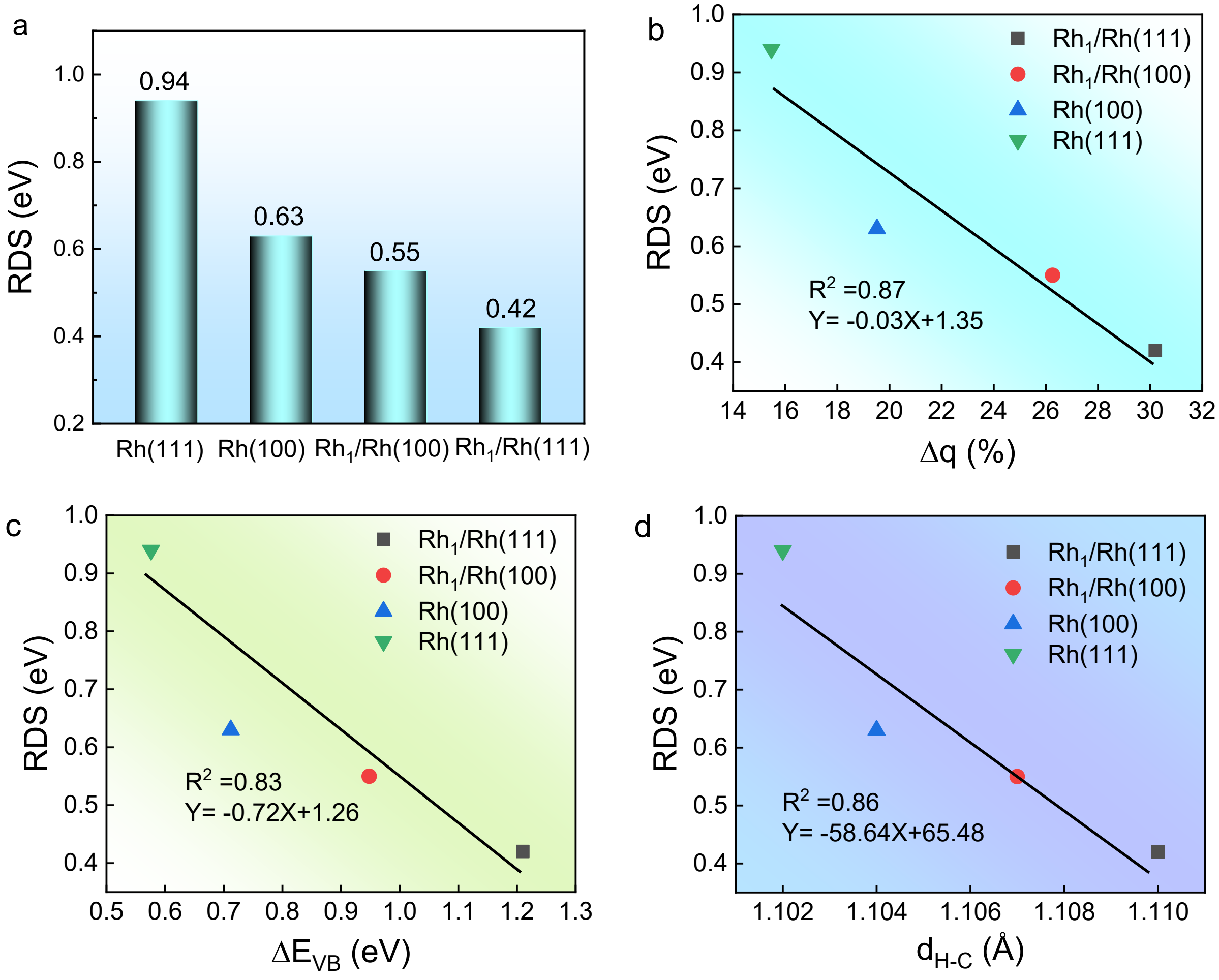

Fig. 8 Atomic CN resolved RDS energy barriers and the correlation between RDS energy barriers and charge gain, H-C length, VB-center shift. (a) Atomic undercoordination lowers the free energy barriers of the rate-determining step for $CH_4$ dehydrogenation. The linear dependence of the RDS energy barriers on (b) charge gain, (c) VB-center shift, and (d) H-C length of adsorbed $CH_4$.

## 5. Conclusions

We have decomposed the ADC for $CH_4$ dehydrogenization into three focusing points. The first one is the nature of the undercoordinated atomic catalyst, the second is catalyst-reactant coupling interaction during physisorption, and the last is the coordination-resolved dehydrogenation energetics and pathways of the atomistic dipolar catalysis. The following summarizes the key findings:

1) The BOLS-NEP notion governs the performance of the atomistic catalyst including adatoms and vacancies. Atomic undercoordination shortens the local bond, densifies and entraps core and bonding electrons, polarizes the valence electrons of the apex or edge atoms to make them $M^P/V^0$ dipoles with excessive charge quantities.

2) The atomistic dipolar $M^P/V^0$ interacts with the reactant molecule by forming the dipolar-substituted hydrogen bond $M^P/V^0$:H-C bond that follows the HBCP regulation initiated for the performance nd functionally of the O:H-O bond.
3) The interplay of $M^P/V^0$:H attraction and $M^P/V^0$:⇔C repulsion lengthens the H-C bond, activating the reactant toward dissociation.
4) The catalyst bond strain, the charge quantity of the $M^P/V^0$, the $\Delta d_{MP:H}$, the $\Delta d_{H\text{-}C}$, and the catalytic capability will be more pronounced in reality as the DFT underestimated the effect of BOLS compared to XPS detection.
5) Modulating the $M^P/V^0$ charge quantity to create the desired electric field should be the unique criterion for efficient ADC.

## Declaration

No conflicting interest is declared.

## Acknowledgment

Financial support from the Guangdong Provincial (No. 2024A1515011094(CQ)) and the National Natural Science Foundation of China (No. 12150100 (BW)) are gratefully acknowledged.

## Appendix 1

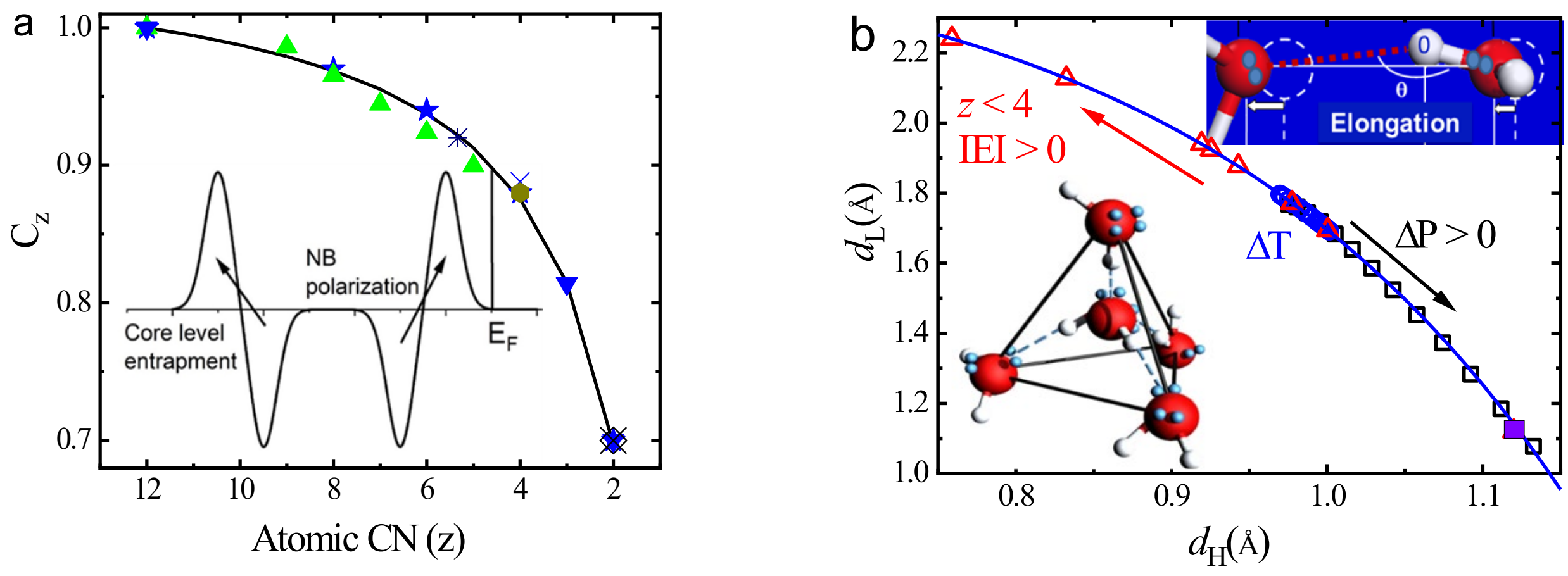

Fig. A1 BOLS-NEP and HBCP notions. (a) The BOLS-NEP notion [9, 39] features the consequence of atomic undercoordination on the local bond contraction, core band self-entrapment, and nonbonding electron polarization (inset). The scattered symbols are experimental results from various sources, see references [9, 39].(Reproduced with Copyright permission from [41] and [9]). There has been plenty of evidence for the BOLS-NEP dictating the permanence of undercoordinated atoms and the size-dependent nanostructures [39]. For instance, 30% bond contraction occurs in monatomic chains and graphene edges [9]. The first layer spacing of an fcc(001) surface contracts by 12% [48]. The skin H-O bond of nanometer-sized water droplets contracts by up to 10% [49] and bond contraction happens mainly in the outermost two atomic layer spacings [49-50]. (b) The HBCP [41-42] notion regulates the coupling O:H-O bond response to a perturbation. The intermolecular O:H nonbond (denoted with subscript L) and the intramolecular H−O bond (subscript H) relax cooperatively under the perturbation of pressure (P), temperature (T), electric filed polarization (E), and molecular undercoordination (z) with inset showing the $H_2O$:$4H_2O$ unit cell and the coupling O:H-O bond.

## Appendix 2 Formulation and XPS derived bond strain

$$\varepsilon = \left(d_z - d_{bulk}\right)/d_{bulk} \qquad (a\text{: strain})$$

$$\Delta E_{VB} = \frac{\int_{E_0}^{E_F} I_z(E)EdE}{\int_{E_0}^{E_F} I_z(E)dE} - \frac{\int_{E_0}^{E_F} I_{bulk}(E)EdE}{\int_{E_0}^{E_F} I_{bulk}(E)dE} \qquad (b\text{: VB center shift})$$

$$E_{VB} = \frac{\int_{E_0}^{E_F} I_z(E)EdE}{\int_{E_0}^{E_F} I_z(E)dE} \qquad (c\text{: VB center})$$

$$\Delta q = \int_{E(I=0)}^{E_F}\left[\frac{I_z(E)}{\int_{E_0}^{E_F} I_z(E)dE} - \frac{I_{bulk}(E)}{\int_{E_0}^{E_F} I_{bulk}(E)dE}\right]dE \qquad (d\text{: charge quantity gain})$$

$$\Delta \text{LDOS} = \frac{I_z(E)}{\int_{E_0}^{E_F} I_z(E)dE} - \frac{I_{bulk}(E)}{\int_{E_0}^{E_F} I_{bulk}(E)dE} \qquad (e\text{: differential LDOS})$$

The spectral intensity I(E) integration in (b) starts from the VB bottom ($E_0$) to the $E_F$, and in (d) the integration over the PDPS peak. Subscript bulk is the reference, and subscript z represents the coordinated catalyst.